\documentclass[letters,usenatbib]{mnras}

\usepackage{newtxtext,newtxmath}

\usepackage[T1]{fontenc}
\usepackage{ae,aecompl}
\usepackage{times}
\usepackage{graphicx}
\usepackage{amsmath}
\usepackage{amssymb}
\usepackage{subcaption}
\usepackage{natbib}
\usepackage{upgreek}

\title[MG~J0414+0534 flux-ratio anomaly]{A flux-ratio anomaly in the CO spectral line emission from gravitationally-lensed quasar MG~J0414+0534}

\author[H.~R.~Stacey and J.~P.~McKean]{
H. R. Stacey$^{1,2}$\thanks{E-mail: h.r.stacey@astro.rug.nl} and J. P. McKean$^{1,2}$\\
$^{1}$ASTRON, Netherlands Institute for Radio Astronomy, Postbus 2, 7990 AA, Dwingeloo, the Netherlands\\
$^{2}$Kapteyn Astronomical Institute, PO Box 800, 9700 AV Groningen, the Netherlands\\}

\date{Accepted 2018 August 16. Received 2018 August 15; in original form 2018 July 27}

\pubyear{2018}

\begin{document}
\label{firstpage}
\pagerange{\pageref{firstpage}--\pageref{lastpage}}
\maketitle

\begin{abstract}
We present an analysis of archival observations with the Atacama Large (sub-)Millimetre Array (ALMA) of the gravitationally lensed quasar MG~J0414+0534, which show four compact images of the quasar and an Einstein ring from the dust associated with the quasar host galaxy. We confirm that the flux-ratio anomalies observed in the mid-infrared and radio persists into the sub-mm for the continuum images of the quasar. We report the detection of CO (11--10) spectral line emission, which traces a region of compact gas around the quasar nucleus. This line emission also shows evidence of a flux-ratio anomaly between the merging lensed images that is consistent with those observed at other wavelengths, suggesting high-excitation CO can also provide a useful probe of substructures that is unaffected by microlensing or dust extinction. However, we do not detect the candidate dusty dwarf galaxy that was previously reported with this dataset, which we conclude is due to a noise artefact. Thus, the cause of the flux-ratio anomaly between the merging lensed images is still unknown. The composite compact and diffuse emission in this system suggest lensed quasar-starbursts will make excellent targets for detecting dark sub-haloes and testing models for dark matter.
\end{abstract}


\section{Introduction}
\label{section:intro}

Almost all well-studied four-imaged gravitationally-lensed quasars have image flux ratios that are inconsistent with those expected from a smooth lensing mass distribution (e.g. \citealt{Koopmans:2003}). These so-called {\it flux-ratio anomalies} are primarily thought to be due to a local perturbation in the mass model that effects one or more of the image magnifications. These perturbations can be in the form of a population of low-mass sub-haloes either within the lensing galaxy or along the line-of-sight that are predicted by dark matter simulations (e.g. \citealt{Mao:1998,Dalal:2002,Xu:2012}), or due to massive companion satellite galaxies of the main lens (e.g. \citealt{McKean:2007,More:2009}), or from unaccounted for mass structure in the form of large-scale galactic discs (e.g. \citealt{Gilman:2017,Hsueh:2018}). In addition, disentangling these flux-ratio anomalies from microlensing by stars within the lensing galaxy, or extrinsic effects such as scintillation or extinction, can be difficult due to the compactness of the lensed images \citep{Sluse:2013}.

The confirmation of a persistent flux-ratio anomaly has typically required high-resolution radio or mid-infrared observations, where the source is expected to be largely immune to microlensing and dust extinction \citep{Minezaki:2009,Jackson:2015}. Recently, the narrow-line emission from gravitationally-lensed quasars has provided an alternate strategy to measure reliable flux ratios, as this emission is expected to probe scales larger than is typically affected by microlensing and is not expected to be effected by intrinsic variability \citep{Moustakas:2003,Nierenberg:2016,Nierenberg:2017}.

However, the flux-ratio anomaly method has thus far been limited by the very small statistical samples of around 7 quadruply-imaged lensed quasars that are suitable for constraining dark matter \citep{Dalal:2002,Xu:2015}. In the near and long-term, large-scale surveys at optical wavelengths through, for example, the Dark Energy Survey (DES; e.g. \citealt{Agnello:2017}) and {\it Euclid} \citep[e.g.][]{Serjeant:2014} will substantially increase the number of known gravitationally-lensed quasars with four images and provide a sample size that is large enough to rule out models for dark matter \citep{Gilman:2018}. These systems will need to be followed-up with either radio/mid-infrared continuum or narrow-line flux measurements to establish robust flux ratios, which themselves may have unknown systematics (for example, on the assumed source size). Therefore, new methods for robustly measuring the flux ratios of the lensed images are required to test for such systematics, provide independent measurements for the same objects, or to increase sample sizes in general.

In this letter, we present the detection of a flux-ratio anomaly in the high-excitation CO (11--10) emission line from the gravitationally-lensed quasar MG~J0414+0534, which reveals a new and independent channel for quantifying low-mass dark matter haloes.

MG~J0414+0534 is a radio-loud quasar at $z=2.64$ that is gravitationally-lensed into four images by a foreground galaxy at $z=0.96$. While the optical and near-infrared fluxes of this lensed quasar are effected by microlensing and extinction, the anomalous fluxes of images A2 and B, relative to image A1, persist at mid-infrared and radio wavelengths. The nature of the flux-ratio anomalies are quite complex; image B is thought to be affected by an optically-luminous companion galaxy to the main lens (object X; \citealt{Falco:1997,Ros:2000,Minezaki:2009}), whereas image A2 is thought to be perturbed by a dark sub-halo with a mass $\sim10^{7}$~M$_{\odot}$ (within the Einstein radius; \citealt{Macleod:2013}).

A possible detection of mm-continuum emission from this optically dark sub-halo (object Y, located 0.6~arcsec East of image A2) was recently reported by \citet{Inoue:2017} from deep Atacama Large (sub-)Millimetre Array (ALMA) observations. If genuine, this would confirm that the flux-ratio anomaly between the two merging images is due to a single low mass sub-halo, as predicted from cold dark matter models.

Here, we re-analyse the ALMA observations of MG~J0414+0534 to investigate the flux-ratio anomaly in the continuum and, for the first time, also in the molecular gas emission from a lensed quasar at mm-wavelengths. We confirm that there is a flux-ratio anomaly in the continuum emission from the quasar, but we do not find any evidence for the candidate dusty dwarf galaxy, object Y. We detect a flux-ratio anomaly in the mm molecular gas emission, which is consistent with that seen at radio and mid-infrared continuum measurements for this object. In Sections~\ref{section:obs} and \ref{section:results} we report the observations and results, respectively, and in Section~\ref{section:discussion} we discuss our results in the context of substructure searches and consider future prospects for a direct detection of a dark sub-halo.

\begin{figure*}
\includegraphics[width=0.49\textwidth]{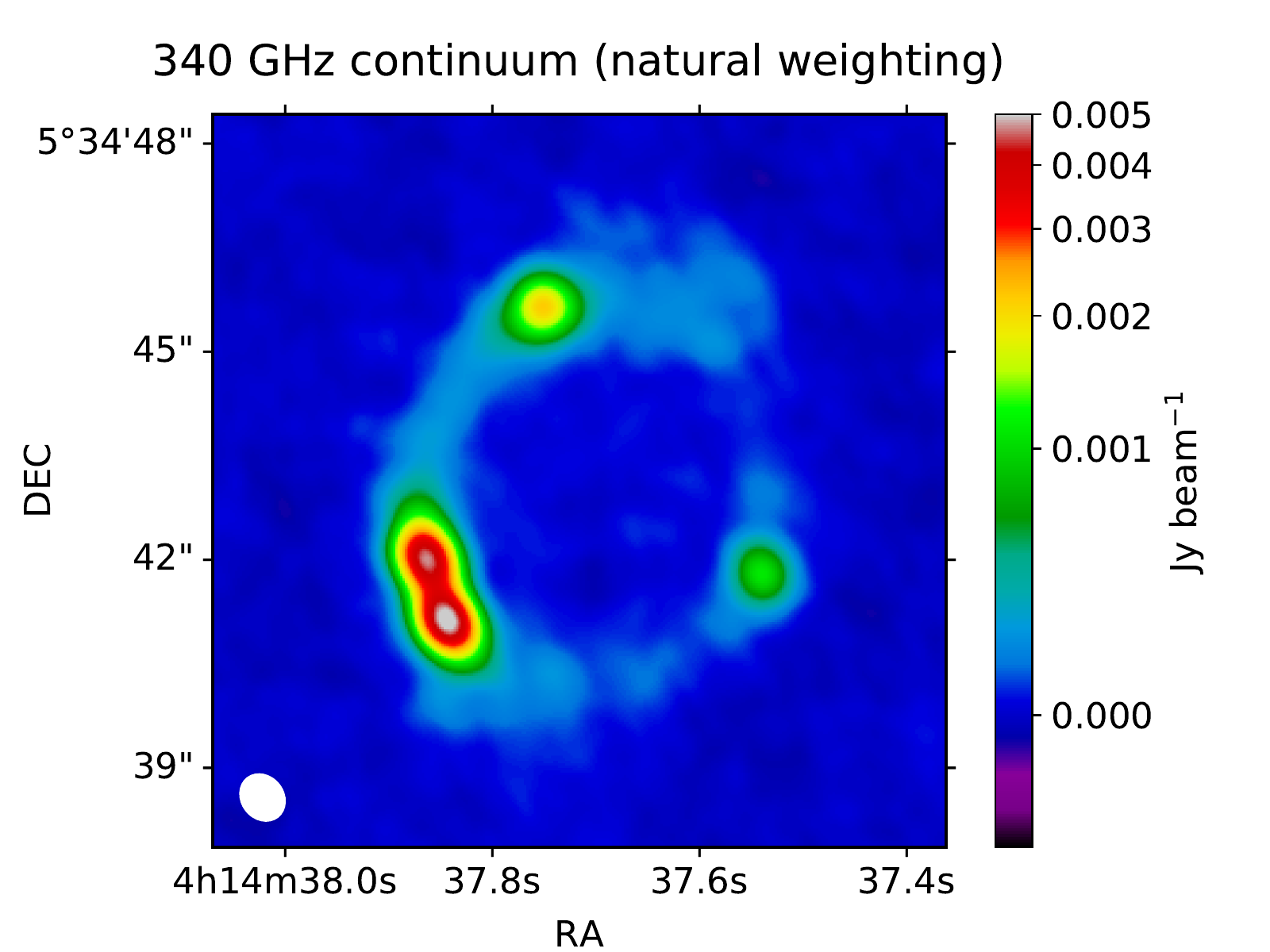}
\includegraphics[width=0.49\textwidth]{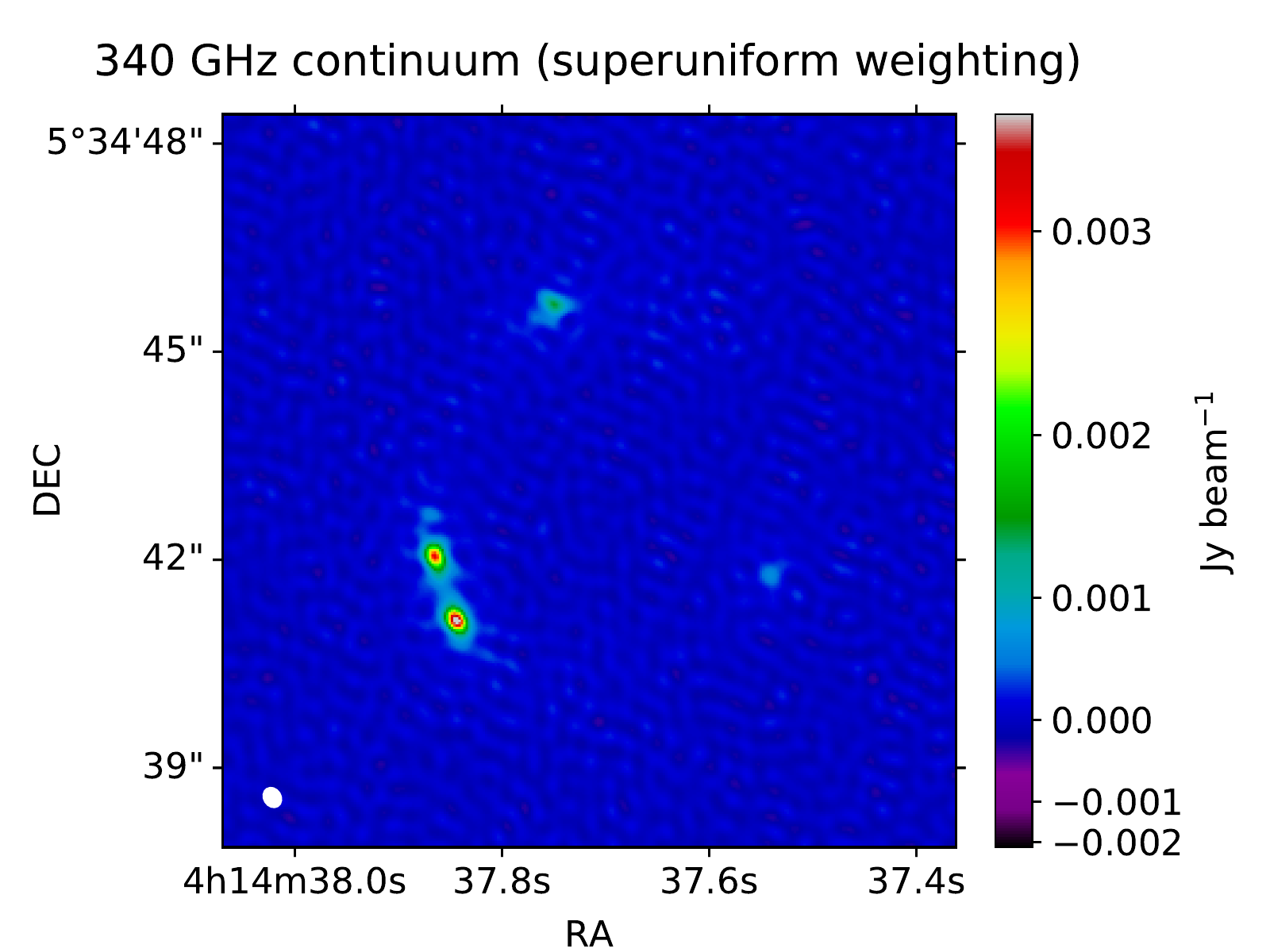}
\caption{ALMA 340~GHz continuum image of MG~J0414+0534 using a natural (left) and a super-uniform (right) weighting of the visibilities. Clockwise from the brightest image A1 are images A2, B and C. The naturally weighted image has a beam-size of $0.30 \times 0.25$~arcsec at a positional angle of 37.6~deg East of North, and an rms map noise of $19~\upmu$Jy~beam$^{-1}$. The super-uniform weighted image has a beam-size of $0.12 \times 0.10$~arcsec at a positional angle of 32.3~deg East of North, and an rms map noise of $91~\upmu$Jy~beam$^{-1}$.}
\label{fig:cont_images}
\end{figure*}

\section{Observations and data reduction}
\label{section:obs}

MG~J0414+0534 was observed with ALMA at a central frequency of 342~GHz (Band 7) on 2015 June 13 and 2015 August 14 (PI Inoue; project code 2013.1.01110.S). This frequency also provided an observation of the CO (11--10) emission line ($\nu_{\rm rest}=1267.014$~GHz). The first set of observations were split into two blocks of about 80~mins duration each, with a combined on-source integration time of about 100 mins.  In total, 35 antennas from the 12-m array were used, with baselines ranging from 21 to 687~m. The second observation (third observing block) was carried out over about 135 mins, with about 50 mins on-source. Here, 42 antennas from the 12~m array were used with baselines between 15 and 1575~m. The flux-density calibration was carried out using J0238+166, and the complex gains (amplitude and phase) as a function of frequency and time were solved for using J0423$-$013 and J0426+0518, with phase referencing carried out at 7~min intervals. For the second set of observations, which included longer baselines, an observation of J0406+0637 was used to correct for the residual delays over the large bandwidth. The data were recorded using both linear polarizations (XX and YY) over four spectral windows centred on 335.0, 336.9, 347.0 and 349~GHz with 2~GHz bandwidth each. The visibilities were correlated using an integration time of 2~s and 128 channels per spectral window.

The raw visibility dataset was processed using the ALMA pipeline within the Common Astronomy Software Applications ({\sc{casa}}) package to produce calibrated visibilities, which were inspected as a function of time, {\it uv}-distance and frequency to check the overall quality of the reduction. As the spectral line is spread across two spectral windows, we modified the pipeline to prevent flagging of the edge channels and allow a bandpass solution to be derived for the whole spectral window. After data inspection and flagging of a bad antenna, imaging and self-calibration of the continuum were carried out within {\sc{casa}} using a super-uniform to natural weighting to determine an accurate source model that accounted for both the small and large-scale structure of the target. Phase-only self-calibration over several time-scales, starting with each scan length, was used to progressively build the model for the source surface brightness distribution, down to a solution interval of 240~s. Note that this phase-only self-calibration was needed to both align the observations that were taken on different days and to correct for residual phase errors that remained after phase referencing. The self-calibrated phase corrections that were derived for the continuum as a function of time were then applied to the spectral line channels. The spectral line data was prepared by fitting a model to the line-free continuum and subtracting this model from the visibilities.

The flux densities of the compact components (continuum and spectral line) were measured by fitting models of delta functions to each image directly in the visibility plane using {\sc UVMultiFit} \citep{Marti-Vidal:2014}, which are presented in Table~\ref{table:line_fluxes}.

\section{Results}
\label{section:results}

\subsection{Continuum}

Using a natural weighting of the visibility data, we detect an Einstein ring of diffuse emission as well as four bright, compact components that we identify as composite thermal dust and non-thermal synchrotron emission, respectively (see Fig.~\ref{fig:cont_images}). This combination is consistent with the overall FIR--radio spectral energy distribution of MG~J0414+0534 \citep{Stacey:2018}. The total flux-density is $S_{\rm 340~GHz} = 24.3\pm1.2$~mJy, of which about 50~percent is contained in the compact images.

Contrary to the findings of \cite{Inoue:2017}, we do not detect a source near image A2 (object Y). We do find a $4\sigma$ surface brightness peak at this position in the pipeline-reduced data before self-calibration, in addition to a number of higher significance artefacts that are caused by residual phase errors; given that the noise is correlated in the sky-plane for interferometric data, the requirements for blind detections in the radio are typically $>6.5\sigma$. These artefacts, including object Y, do not remain in the data after subsequent steps of self-calibration. The rms noise in our naturally-weighted image, after correcting for the residual antenna-based phase errors is $19~\upmu$Jy~beam$^{-1}$, in comparison to the $65~\upmu$Jy~beam$^{-1}$ from the analysis by \citet{Inoue:2017}. To test that we did not bias the result with our choice of model during self-calibration, we also used an initial model that included object Y (as described in Section~\ref{section:obs}). However, the calibration does not converge to a solution were object Y remains after the process. Therefore, we conclude that object Y is a noise artefact resulting from residual phase errors rather than a genuine astronomical source.

\begin{figure}
\includegraphics[width=0.49\textwidth]{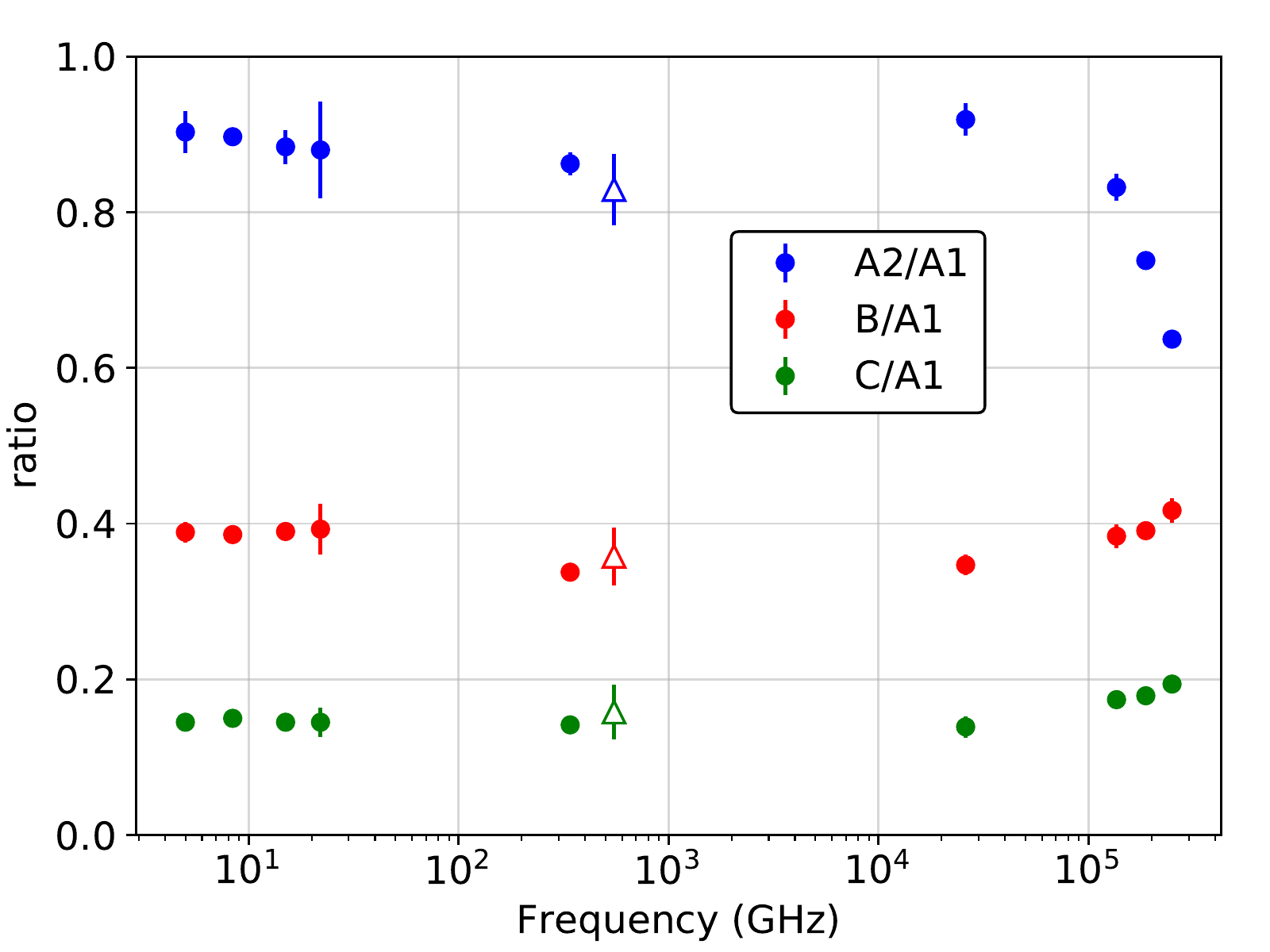}
\caption{The continuum flux ratio of images A2, B and C relative to image A1 from the radio to near-infrared. The CO (11--10) flux ratios are also shown, marked with open triangles at an arbitrary frequency. The flux ratios measured in the near-infrared are affected by extinction and microlensing. The radio, mid-infrared and near-infrared measurements are taken from \citet{Katz:1997}, \citet{Macleod:2013}, and \citet{Kochanek:1999}, respectively.}
\label{fig:all_ratios}
\end{figure}

With super-uniform weighting of the visibility data, we detect four compact components that are coincident with the optical and radio core of the AGN (see Fig.~\ref{fig:cont_images}). The merging images are also marginally resolved and show evidence for an additional low surface brightness component, likely due to some contribution from compact dust emission. We fit a model of delta functions to the lensed images in the visibility plane (described in Section~\ref{section:obs}) for which we use only epoch 2 due to its higher angular resolution. We also attempted to fit Gaussian models, however, as the images are only marginally resolved and there is some evidence of extended dust emission also, the fitting did not produce stable results. We detect a flux-ratio anomaly between images A1 and A2 (see Table~\ref{table:flux_ratios}), as seen at longer wavelengths, which have a ratio of $S_{\rm A2}/S_{\rm A1} = 0.86\pm0.01$ in the continuum. This ratio is more extreme than the flux ratio measured at $\sim$GHz radio frequencies and in the mid-infrared, all of which are dominated by AGN emission. The flux ratios from the radio to near-infrared are shown in Fig.~\ref{fig:all_ratios}. A slight trend can be seen from 1.4 to 340~GHz towards a more extreme flux-ratio anomaly between images A1 and A2. If genuine, this trend is likely the effect of a change in the source structure, where the lensed emission is more extended at lower frequencies due to an additional contribution from the steep-spectrum jet emission and the larger size of the synchrotron self-absorbed AGN core. Similarly, a less extreme flux ratio in the mid-infrared may be explained by the dusty torus being extended on scales larger than the sub-mm core.

\subsection{CO (11--10)}

\begin{figure}
\centering
\includegraphics[width=0.47\textwidth]{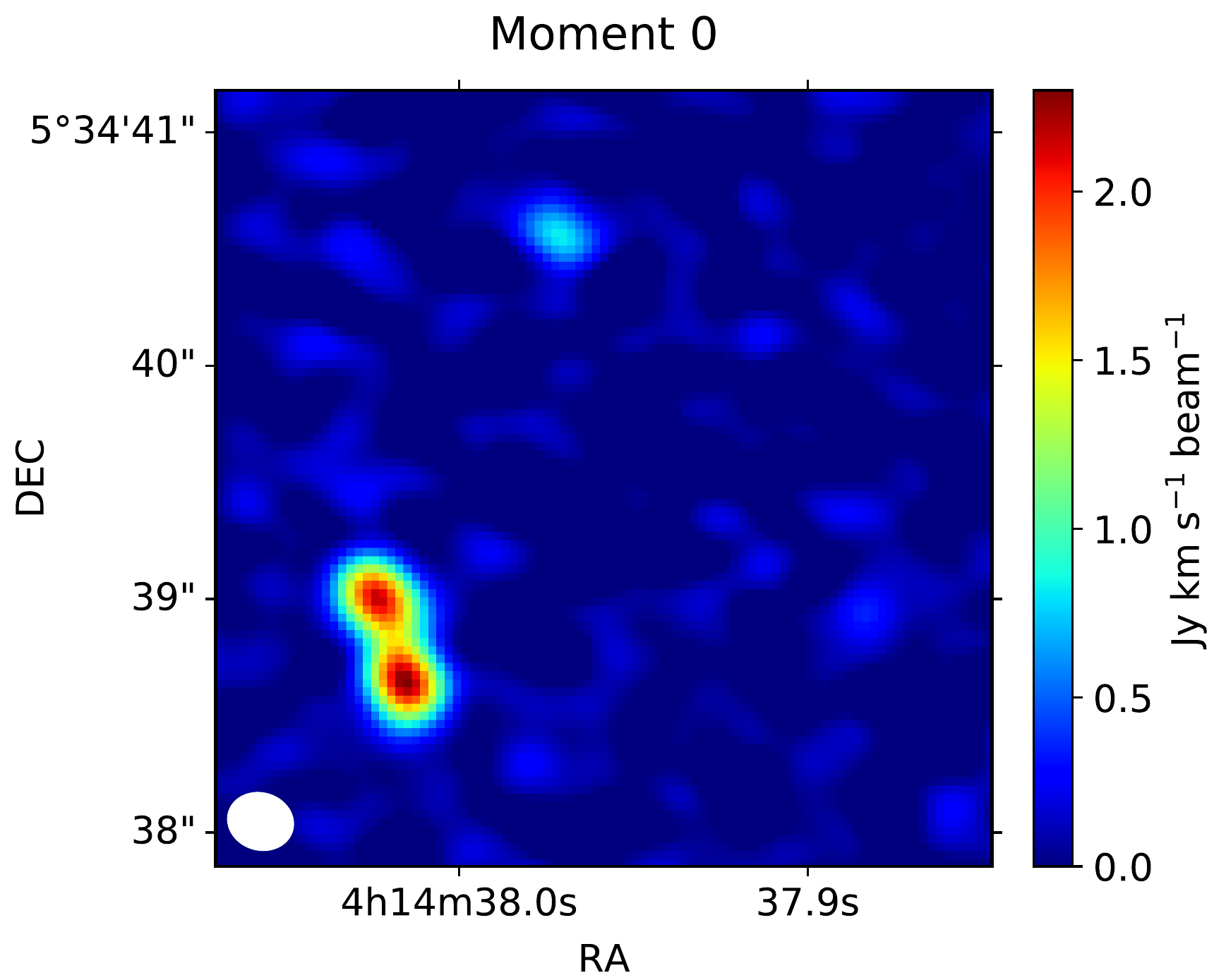}
\caption{The CO (11--10) integrated line intensity (moment-zero) using a Briggs weighting scheme for the visibility data.}
\label{fig:mom0}
\end{figure}

\begin{figure}
\centering
\includegraphics[width=0.48\textwidth]{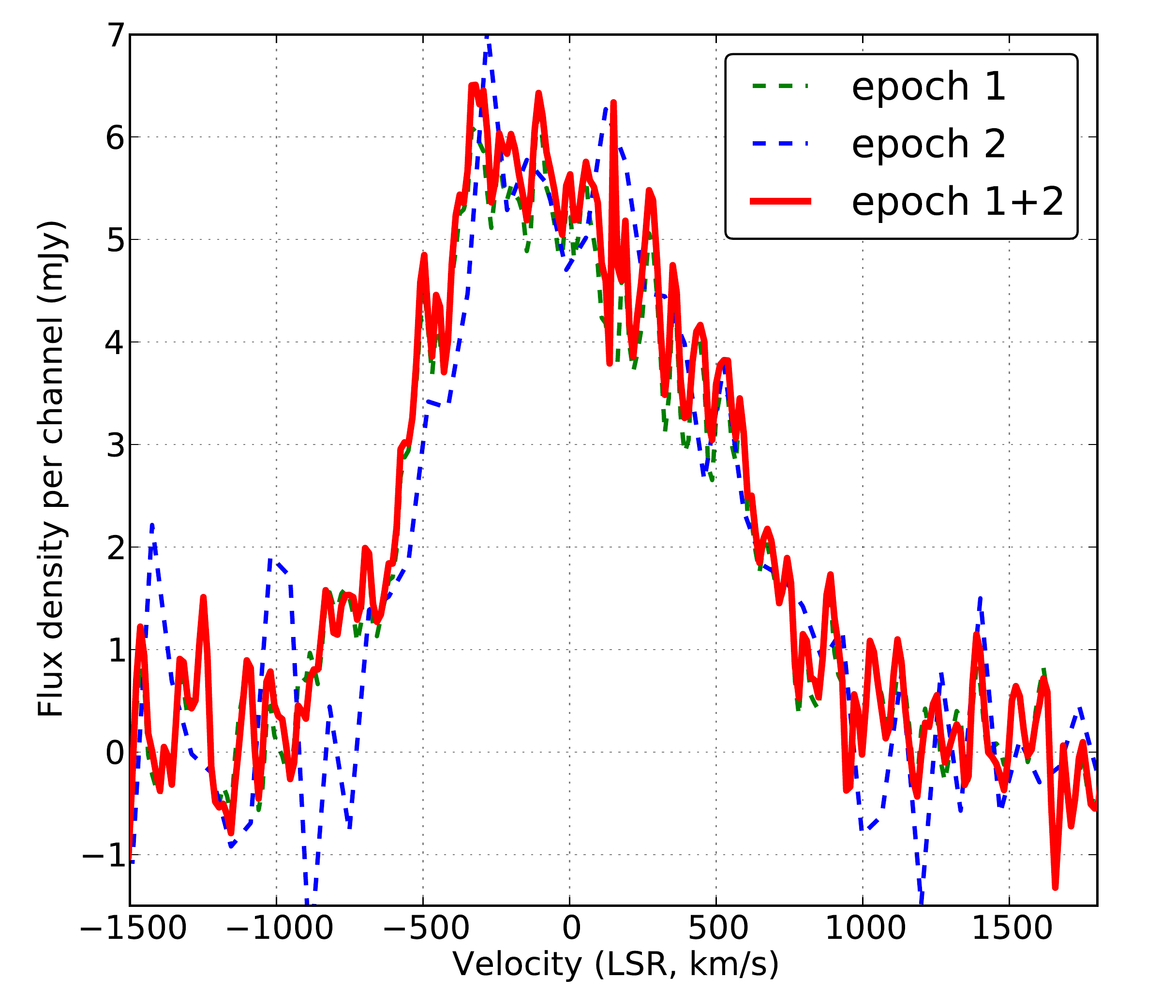}
\caption{The CO (11--10) line profile from MG~J0414+0534 for images A1 and A2 (red). The dotted lines show the spectra from 2015 June 13 (green) and 2015 August 14 (blue), and demonstrate that there is no change in the line profile over this timescale. The data for epoch 2 has been smoothed with a boxcar kernel of width 5 channels. The systemic velocity is relative to $z = 2.639$, using the radio definition of the velocity.}
\label{fig:line_profile}
\end{figure}

\begin{figure}
\begin{subfigure}[b]{0.51\textwidth}
\centering
\includegraphics[width=\textwidth]{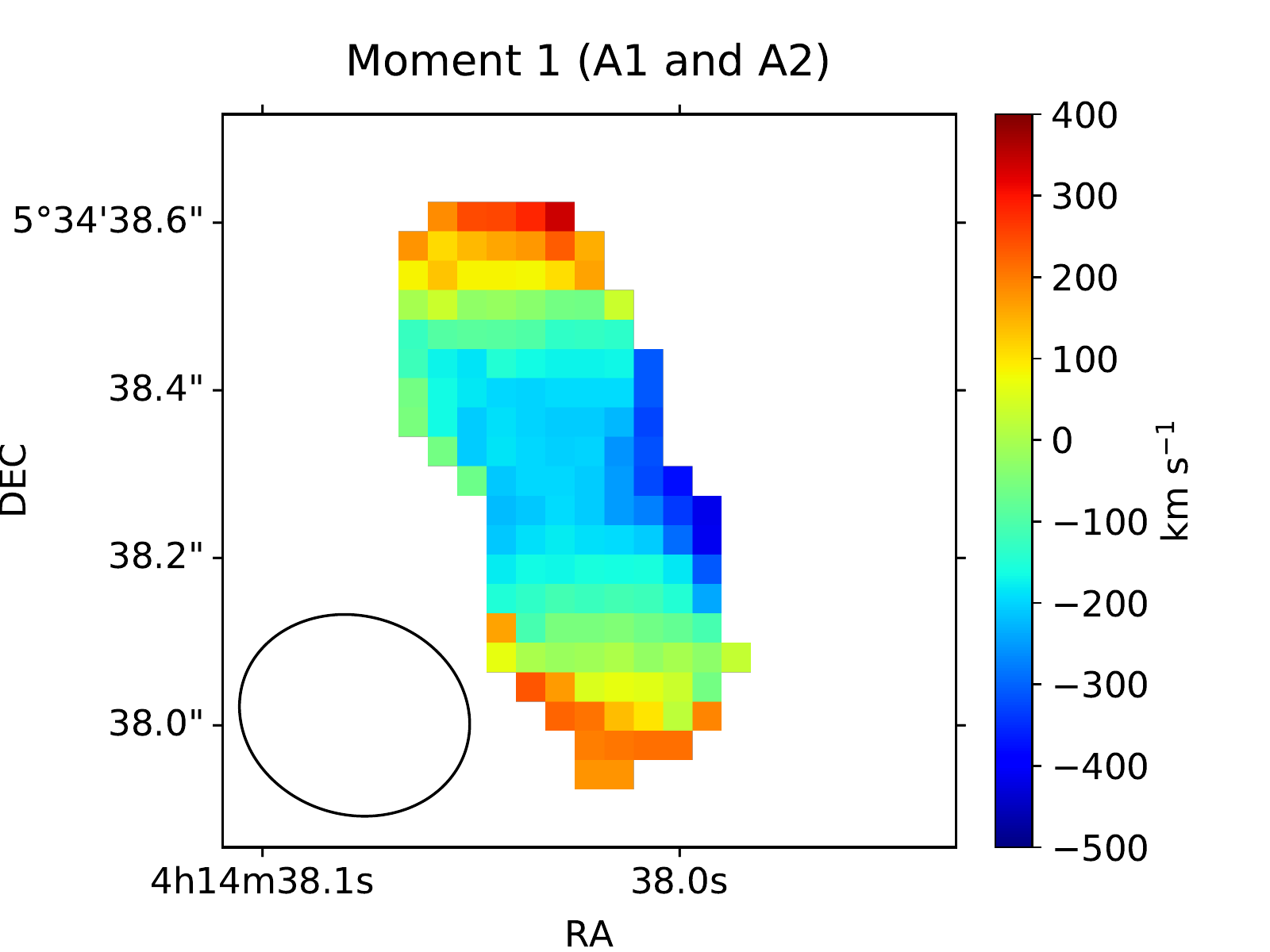}
\end{subfigure}
\begin{subfigure}[b]{0.51\textwidth}
\centering
\includegraphics[width=\textwidth]{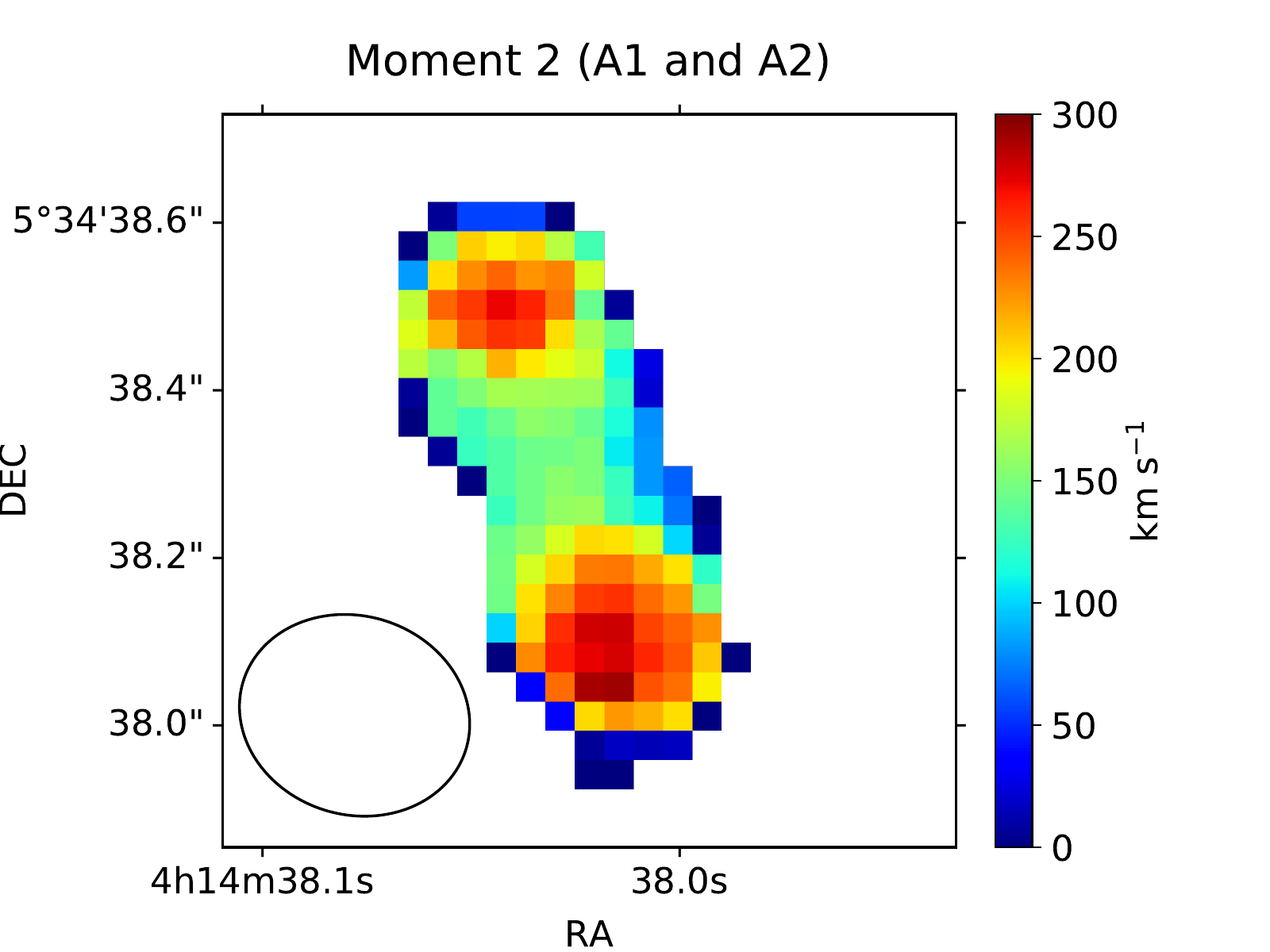}
\end{subfigure}
\caption{The CO (11--10) velocity-field (moment one; upper) and velocity dispersion (moment two; lower), for the merging images A1 and A2. Pixels below the $5\sigma$-level are masked out. The synthesized beam FWHM is shown in the bottom left-hand corner.}
\label{fig:moment_maps}
\end{figure}

The CO (11--10) molecular gas emission from MG~J0414+0534 is clearly detected in the moment-zero map (see Fig.~\ref{fig:mom0}), which is found to be coincident with the four images of the lensed quasar. From extracting the one-dimensional spectral profile over the merging images A1 and A2 (see Fig.~\ref{fig:line_profile}), we find that the line profile is broad and asymmetric, with a higher flux density in the blue-shifted component. The spatially resolved velocity structure for images A1 and A2 (see Fig.~\ref{fig:moment_maps}) shows that there is evidence of a rotating gas structure around the central black hole of the quasar; also, the blue-shifted component approaches the critical curve of the lens, and therefore subject to a higher magnification. Thus, the asymmetric line profile we observe is due to differential magnification across the velocity field, rather than an intrinsic property of the source, as has been seen in the molecular gas emission from other lensed galaxies \citep{Rybak:2015,Paraficz:2018}. A single Gaussian fit to the line profile from images A1 and A2, gives a line FWHM of $1080\pm20$~km\,s$^{-1}$. The gas also has a high velocity dispersion of 300~km\,s$^{-1}$ at the position of the black hole, as seen in the moment-two map of images A1 and A2 (see Fig.~\ref{fig:moment_maps}).

We find that the line emission from the merging images is marginally spatially-resolved. The compact size, high velocity dispersion and broad profile of the line emission are consistent with warm, dense gas in the vicinity of the AGN. While large velocities are found in the broad line region of AGN, \cite{Lawrence:1995} find Balmer lines with FWHM $\sim4000$~km\,s$^{-1}$ for this quasar, suggesting the gas is outside this region. There is no evidence of a change in the line profile between the two epochs (2015 June and August) that would suggest the gas is intrinsically variable or subject to microlensing.

We fit models to the lensed images in the visibility plane (described in Section~\ref{section:obs}), using the channels around the peak of the line in the velocity range $-350$ to 0~km\,s$^{-1}$. We select this range of the data as it has the highest signal-to-noise ratio and the emission is compact enough that it can be well approximated by delta functions. Furthermore, this corresponds to the blue-shifted component of the line emission that is close to the critical curve, and therefore should have a flux ratio that is close to unity for a smooth mass distribution in the absence of substructure. The measured flux ratios of the line emission are given in Table~\ref{table:flux_ratios}. 

The flux ratios in the CO between images A1 and B, and between images C and B, are consistent with the anomalies detected in the radio and mid-infrared bands, which are attributed to the secondary lensing galaxy (object X). We detect a flux-ratio anomaly between the merging images of $S_{\rm A2}/S_{\rm A1} = 0.83\pm0.05$. Image A2 is suppressed relative to A1, as observed at mid-infrared \citep{Minezaki:2009,Macleod:2013} and radio wavelengths \citep[for example]{Katz:1997}. The similarity of the flux ratios of images B and C in the CO and mid-infrared suggests that the gas is on a scale comparable to the AGN torus, as primarily observed in the mid-infrared. However, the flux-ratio anomaly between images A1 and A2 is more extreme than observed in the mid-infrared or mm-continuum as the blue-shifted component of the line emission is likely offset from the position of the central black hole and, being close to the critical curve, is also preferentially magnified. This is the first detection of a flux-ratio anomaly in the mm-line emission from a gravitationally lensed quasar.

\begin{table}
\begin{center}
\caption{The flux densities of the 340~GHz continuum and CO (11--10) line emission (measured between $-300$ and 0~km\,s$^{-1}$).}
\begin{tabular*}{0.29\textwidth}{p{1cm} | p{1.5cm} p{1.5cm} }
Image & $S_{\rm cont}$~(mJy) &  $S_{\rm line}$~(mJy) \\ \hline
A1 & $4.43\pm0.05$ & $2.54\pm0.09$ \\
A2 & $3.82\pm0.05$ & $2.11\pm0.09$ \\
B & $1.50\pm0.05$ & $0.91\pm0.09$ \\
C & $0.63\pm0.05$ & $0.40\pm0.09$ \\ \hline
\end{tabular*}
\label{table:line_fluxes}
\end{center}
\end{table}

\begin{table}
\begin{center}
\caption{The flux ratios of the lensed images at 340~GHz continuum and CO (11--10) line emission (measured between $-300$ and 0~km\,s$^{-1}$).}
\begin{tabular*}{0.29\textwidth}{p{1cm} | p{1.5cm} p{1.5cm} }
Images & $R_{\rm cont}$ & $R_{\rm line}$ \\ \hline
A2/A1 & $0.86\pm0.01$ & $0.83\pm0.05$ \\
B/A1 & $0.34\pm0.01$ & $0.36\pm0.04$ \\
C/A1 & $0.14\pm0.01$ & $0.16\pm0.04$ \\ \hline
\end{tabular*}
\label{table:flux_ratios}
\end{center}
\end{table}

\section{Discussion and Conclusions}
\label{section:discussion}

Although the image flux ratios of gravitationally lensed quasars can be used to test models of galaxy formation and dark matter, the technique is currently limited by the small number of suitable objects that can be used in the analysis (e.g. \citealt{Xu:2015}). Here, we have shown the detection of a flux-ratio anomaly in the molecular gas from a lensed quasar for the first time. By targeting the high excitation CO (11--10) emission, we are able to probe a compact region that is close to the central super-massive black hole, which gives an independent measurement to the flux ratios observed in the mm-continuum and other wavebands. In addition, we have shown that there is no evidence of variability in the line emission from either intrinsic or extrinsic effects and, as the dust is optically thin to the line emission, the CO (11-10) flux-ratios are not affected by extinction. Thus, the study of molecular line emission appears to be a promising method to measure flux-ratio anomalies.

As well as providing an independent measurement of the flux ratios, studies in the mm-regime have several advantages over other wavelengths. First, as shown here and in many other cases, the thermal dust emission in high redshift objects is sufficiently extended to produce an Einstein ring; such data can provide important constraints to the lensing  macro-model, even at a low-angular resolution (e.g. \citealt{Rybak:2015a}). Therefore, lenses that produce only two-images could also now be used in the analysis, increasing the sample size. Second, interferometric arrays such as ALMA can provide sufficiently high angular resolution ($>10$~mas) such that any degeneracies between the inferred level of substructure and the source model (often assumed to be point-like) can be minimised. Finally, 70~percent of known gravitationally lensed quasars are FIR-bright \citep{Stacey:2018}, and so are expected to have detectable high excitation gas associated with the quasar. This, coupled with the large-scale lens surveys being carried out with DES, for example, will provide a large population of potential targets for this work in the near-term. A pilot study of a well-selected sample of lensed quasars to investigate this is planned.

Our study of MG~J0414+0534 has not revealed the cause of the flux-ratio anomaly, although the similar flux ratios observed here and at other wavebands does suggest that it is due to some perturbation in the lensing potential. However, we have demonstrated that it is not due to a candidate dwarf satellite close to images A1 and A2 that was previously reported. As the magnitude of the anomaly is consistent with other studies, we have not tested the possible mass models that are consistent with the data, which we defer to a follow-up paper. This will include combining the continuum and line emission data (e.g. \citealt{Hezaveh:2013}) to determine whether the perturbation is due to a sub-halo or a more massive structure that is currently unaccounted for in the lensing macro-model.

\section*{Acknowledgements}

We thank Leon Koopmans and the anonymous referee for their comments. This paper makes use of the following ALMA data: ADS/JAO.ALMA\#2013.1.01110.S. ALMA is a partnership of ESO (representing its member states), NSF (USA) and NINS (Japan), together with NRC (Canada) and NSC and ASIAA (Taiwan) and KASI (Republic of Korea), in cooperation with the Republic of Chile. The Joint ALMA Observatory is operated by ESO, AUI/NRAO and NAOJ. This research made use of Astropy, a community-developed core Python package for Astronomy \citep{Astropy:2018}.

\bibliographystyle{mnras}
\bibliography{references}

\bsp
\label{lastpage}
\end{document}